\documentclass[]{article}
\usepackage[USenglish]{babel}
\usepackage[utf8]{inputenc}
\usepackage[T1]{fontenc}
\usepackage{spconf}
\usepackage{relsize,algorithm,algpseudocode,amsmath,amssymb,amsthm,bbm,cite,color,enumitem,flushend,graphicx,microtype,setspace,url}
\usepackage{tikz}

\usepackage{csquotes}
\usepackage{subfig}
\usepackage[figurename=Fig.,font=small]{caption}
\usepackage{epstopdf}
\usepackage{verbatim}

\usepackage{pgfplots}
\pgfplotsset{compat=newest}
\usetikzlibrary{external,calc,positioning,automata,shapes,decorations.pathreplacing,decorations.text}
\usetikzlibrary{plotmarks}
\usetikzlibrary{arrows.meta}
\usepgfplotslibrary{patchplots}
\usepackage{grffile}
\usepackage{amsmath}

\pgfplotsset{plot coordinates/math parser=false}

\pgfplotsset{
select coords between index/.style 2 args={
    x filter/.code={
        \ifnum\coordindex<#1\def\pgfmathresult{}\fi
        \ifnum\coordindex>#2\def\pgfmathresult{}\fi
    }
},
vijaya/.style n args={4}{
    color=white!75!cyan,line width=5pt,
    x filter/.code={
        \pgfmathparse{\pgfmathresult+#3*cos(#4)}
        \ifnum\coordindex<#1\def\pgfmathresult{}\fi
        \ifnum\coordindex>#2\def\pgfmathresult{}\fi
    },
    y filter/.code={
        \pgfmathparse{\pgfmathresult+#3*sin(#4)}
    }
},
tissa/.style n args={5}{
    draw=none,
    postaction={decorate,decoration={text effects along path,
    text={#1},raise=#2,
    text align=center,
    text effects/.cd, 
    text along path,
    every character/.style={scale=#3},
    }},
    select coords between index={#4}{#5},
}}

\newlength\figureheight
\newlength\figurewidth

\usepackage[hidelinks]{hyperref}
\usepackage[style=long,numberline,savewrites=true,acronym,nomain,hyperfirst=false]{glossaries}

{}
{}
{}
{}
{}
{}
{}

\newcommand{\h}{\mathbf{h}}

\newcommand{\p}{\mathbf{p}}
\newcommand{\q}{\mathbf{q}}

\newcommand{\x}{\mathbf{x}}

\newcommand{\I}{\mathbf{I}}

\newcommand{\N}{\mathbf{N}}

\newcommand{\R}{\mathbf{R}}

\newcommand{\Y}{\mathbf{Y}}
\newcommand{\Z}{\mathbf{Z}}

\newcommand{\setC}{\mathcal{C}}

\newcommand{\setN}{\mathcal{N}}

\newcommand{\setQ}{\mathcal{Q}}

\newcommand{\Real}{\mbox{$\mathbb{R}$}}
\newcommand{\Compl}{\mbox{$\mathbb{C}$}}

\newcommand{\rmq}{\mathrm{q}}

\newcommand{\argmin}{\operatornamewithlimits{argmin}}

\newcommand{\Exp}{\mathbb{E}}
\newcommand{\norm}[1]{\left\lVert#1\right\rVert}
\newcommand{\herm}{\mathrm{H}}
\renewcommand{\Im}{\mathrm{Im}}
\renewcommand{\Pr}{\mathbb{P}}

\renewcommand{\Re}{\mathrm{Re}}
\newcommand{\sgn}{\mathrm{sgn}}

\newcommand{\tran}{\mathrm{T}}

\newcommand{\red}[1]{\textcolor{red}{#1}}

\newcommand{\rmp}{\mathrm{p}}

\def\showguides{0}

\let\gls\cgls
\let\glspl\cglspl

\glsenableentrycount

\newglossaryentry{gradmse}
           {name=\ensuremath{\Delta_\text{MSE}},
           description={pilot estimation mean squared error difference}}
           
\newglossaryentry{ind}
           {name=\ensuremath{\mathbbm{1}},
           description={indicator function}}

\newglossaryentry{gradser}
           {name=\ensuremath{\Delta_\text{SER}},
           description={data detection symbol error rate gradient}}

\newglossaryentry{pilotmse}
           {name=\ensuremath{\text{MSE}},
           description={pilot estimation MSE wrt original constellation}}

\newglossaryentry{pilotse}
           {name=\ensuremath{\text{SE}},
           description={pilot estimation SE wrt original constellation}}

\newglossaryentry{dataser}
           {name=\ensuremath{\text{SER}},
           description={data detection $\text{SER}$ wrt expected constellation}}

\newglossaryentry{plesterr}
                 {name=\ensuremath{\rho_{\textnormal{\tiny{offset}}}},
                  description={power offset}}

\newglossaryentry{plesterrest}
                 {name=\ensuremath{\hat{\rho_{\textnormal{\tiny{offset}}}}},
                  description={estimated power offset}}

\newglossaryentry{plesterrestvar}
                 {name=\ensuremath{\sigma^2_{\hat{\rho_{\textnormal{\tiny{offset}}}}}},
                  description={estimated path loss estimation offset}}

\newacronym{3gpp}
           {3GPP}
           {Third-Generation Partnership Project}

\newacronym{dpc}
           {DPC}
           {differential power control}

\newacronym{mdd}
           {MDD}
           {minimum distance detection}

\newacronym{mmse}
           {MMSE}
           {minimum mean squared error}

\newacronym{zf}
           {ZF}
           {zero forcing}

\newacronym{ssb}
           {SSB}
           {synchronization signal block}

\newacronym{dmrs}
           {DMRS}
           {demodulation reference signal}

\newacronym{srs}
           {SRS}
           {sounding reference signal}

\newacronym{prach}
           {PRACH}
           {physical random access channel}

\newacronym{rrh}
           {RRH}
           {remote radio head}

\newacronym{dft}
           {DFT}
           {discrete Fourier transform}

\newacronym{ote}
           {OTE}
           {one time entry}

\newacronym{iid}
           {i.i.d.}
           {independent and identically distributed}

\newacronym{bbu}
           {BBU}
           {baseband unit}

\newacronym{mimo}
           {MIMO}
           {multiple-input multiple-output}

\newacronym{simo}
           {SIMO}
           {single-input multiple-output}

\newacronym{adc}
           {ADC}
           {analog-to-digital converter}

\newacronym{dac}
           {DAC}
           {digital-to-analog converter}

\newacronym{ofdm}
           {OFDM}
           {orthogonal frequency division multiplexing}

\newacronym{csi}
           {CSI}
           {channel state information}

\newacronym{ris}
           {RIS}
           {reflective intelligent surfaces}

\newacronym{comp}
           {CoMP}
           {coordinated multipoint}

\newacronym{bs}
           {BS}
           {base station}

\newacronym{pc}
           {PC}
           {power control}

\newacronym{su}
           {SU}
           {single user}

\newacronym{mu}
           {MU}
           {multi-user}

\newacronym{awgn}
           {AWGN}
           {additive white Gaussian noise}

\newacronym{snr}%
           {\ensuremath{\text{SNR}}}
           {signal-to-noise ratio}

\newacronym{sir}%
           {SIR}
           {signal-to-interference ratio}

\newacronym{rar}
           {RAR}
           {random-access response}
           
\newacronym{rf}
           {RF}
           {radio frequency}

\newacronym{sinr}
           {SINR}
           {signal-to-interference-plus-noise ratio}

\newacronym{se}
           {\ensuremath{\text{SE}}}
           {squared error}

\newacronym{mse}
           {\ensuremath{\text{MSE}}}
           {mean squared error}

\newglossaryentry{stepsize}{
           name={\ensuremath{\delta\hspace{-.15em}\rho}},
           description={step size}}

\newglossaryentry{rho2ser}{
           name={\ensuremath{h}},
           description={$\rho$ to $\text{SER}$ map}}

\newglossaryentry{rho2mse}{
           name={\ensuremath{g}},
           description={$\rho$ to $\text{MSE}$ map}}

\newglossaryentry{plesterresterrstd}{
           name={\ensuremath{\sigma_{|\hat{\rho_\text{\tiny offset}} - \rho_\text{\tiny offset}|}}},
           description={standard deviation of the error of estimating the power offset}}

\newglossaryentry{plesterresterrvar}{
           name={\ensuremath{\Exp\left[|\hat{\rho}_\text{\tiny offset} - \rho_\text{\tiny offset}|^2\right]}},
           description={variance of the error of estimating the power offset}}

\newglossaryentry{compositepower}{
           name={\ensuremath{{\rho}^{\tiny \mathrm{comp}}}},
           description={composite transmit power vector}}

\newglossaryentry{compositepowervec}{
           name={\ensuremath{\boldsymbol{{\rho}}^{\tiny \mathrm{comp}}}},
           description={composite transmit power vector}}

\newglossaryentry{gap}{
           name={\ensuremath{\Delta}},
           description={transmit SNR gap in composite pilot}}

\newglossaryentry{rhostart}{
           name={\ensuremath{\rho_{\mathrm{start}}}},
           description={transmit SNR shift for a ramping stage}}

\newacronym{ser}
           {\ensuremath{\text{SER}}}
           {symbol error rate}

\newacronym{evm}
           {\ensuremath{\text{EVM}}}
           {error vector magnitude}

\newacronym{fer}
           {\ensuremath{\text{FER}}}
           {frame error rate}

\newacronym{tpc}
           {TPC}
           {transmission power control}

\newacronym{qam}
           {QAM}
           {quadrature amplitude modulation}

\newacronym{psk}
           {PSK}
           {phase shift keying}

\newacronym{qpsk}
           {QPSK}
           {quadrature phase shift keying}

\newacronym{ul}
           {UL}
           {uplink}

\newacronym{ue}
           {UE}
           {user equipment}

\newacronym{ls}
           {LS}
           {least-squares}

\newacronym{sls}
           {SLS}
           {scaled least-squares}

\newacronym{blm}
           {BLM}
           {Bussgang linear MMSE}

\newacronym{mrc}
           {MRC}
           {maximum ratio combining}

\newacronym{papr}
           {PAPR}
           {peak-to-average power ratio}

\newacronym{nr}
           {NR}
           {new radio}

\newacronym{is-95}
           {IS-95}
           {interim standard 95}

\newacronym{1g}
           {1G}
           {first generation}

\newacronym{2g}
           {2G}
           {second generation}

\newacronym{3g}
           {3G}
           {third generation}

\newacronym{4g}
           {4G}
           {fourth generation}

\newacronym{5g}
           {5G}
           {fifth generation}

\newacronym{umts}
           {UMTS}
           {unversal mobile telecommunication system}

\newacronym{lte}
           {LTE}
           {Long-Term Evolution}

\newacronym{ecc}
           {ECC}
           {error control coding}

\tikzset{
short/.style={draw,rectangle,text height=3pt,text depth=13pt,
  text width=7pt,align=center,fill=gray!30},
long/.style={short,text width=1.5cm}
}

\title{Pilot-Based Uplink Power Control in Single-UE Massive MIMO Systems With 1-Bit ADCs}
\name{Amila Ravinath, Bikshapathi Gouda, Italo Atzeni, and Antti Tölli\thanks{This work is supported by the Academy of Finland (318927 6G~Flagship, 336449 Profi6, 348396 HIGH-6G, and 357504 EETCAMD) and by the European Commission (101095759 Hexa-X-II).}}
\address{Centre for Wireless Communications, University of Oulu, Finland \\
Emails: \{amila.ravinath, bikshapathi.gouda, italo.atzeni, antti.tolli\}@oulu.fi}

\begin{document}
\setlength{\abovedisplayskip}{-.1mm}
\setlength{\belowdisplayskip}{-.1mm}
\setlength{\abovedisplayshortskip}{.1pt}
\setlength{\belowdisplayshortskip}{.1pt}
\ninept
\maketitle

\begin{abstract}
We propose uplink \gls{pc} methods for massive multiple-input multiple-output systems with 1-bit analog-to-digital converters, which are specifically tailored to address the non-monotonic data detection performance with respect to the transmit power of the \gls{ue}. Considering a single \gls{ue}, we design a multi-amplitude pilot sequence to capture the aforementioned non-monotonicity, which is utilized at the base station to derive \gls{ue} transmit power adjustments via single-shot or \gls{dpc} techniques. Both methods enable closed-loop uplink \gls{pc} using different feedback approaches. The single-shot method employs one-time multi-bit feedback, while the \gls{dpc} method relies on continuous adjustments with 1-bit feedback. Numerical results demonstrate the superiority of the proposed schemes over conventional closed-loop uplink \gls{pc} techniques. 

\textbf{\textit{Index Terms}}---1-bit ADCs, massive MIMO, uplink PC.
\end{abstract}

\glsresetall

\vspace{-4mm}
\section{Introduction}
\vspace{-4mm}

Future wireless communication systems will push the carrier frequencies towards the (sub-)THz region to meet the demand for higher data rates and lower latency \cite{Raj20}. Beamforming with massive \gls{mimo} arrays can compensate for the increased pathloss at (sub-)THz frequencies, allowing for a cell radius comparable to that of sub-6~GHz systems. Fully digital massive \gls{mimo} architectures enable flexible beamforming and large-scale spatial multiplexing \cite{Mar10}, although each antenna requires a dedicated radio-frequency chain. In this setting, the power consumption of the \glspl{adc} scales linearly with the sampling rate and exponentially with the number of quantization bits \cite{Mo15,Li17,Jac17,Atz21b}. Consequently, low-resolution \glspl{adc} have been regarded as a promising enabler for high-frequency massive \gls{mimo} in recent years. In particular, 1-bit \glspl{adc} exhibit the lowest power consumption with a modest performance loss compared to unquantized systems. Several works have focused on the performance analysis of uplink massive \gls{mimo} systems with 1-bit \glspl{adc}, e.g., \cite{Mo15,Li17,Mol17,Ucu18,Atz22,Atz21a}. Importantly, the performance of channel estimation and data detection with 1-bit \glspl{adc} is strongly affected by the operating \gls{snr} and, thus, by the transmit power of the \gls{ue} \cite{Jac17,Atz22,Atz21a}.

Uplink \gls{pc} is crucial for maintaining the \gls{sinr} or the \gls{ser} of each \gls{ue} at the desired level, in addition to guaranteeing energy efficiency. In this regard, cellular networks require both open- and closed-loop uplink \gls{pc} techniques. To this end, a framework for uplink \gls{pc} is introduced in \cite{Yat95} while the uplink \gls{pc} implementation of the \gls{3gpp} \gls{lte} standard is evaluated in \cite{Sim08}. When considering a conventional unquantized massive \gls{mimo} system, increasing the \gls{ue} transmit power always leads to a monotonic improvement in the \gls{ser} performance. However, this is not the case when 1-bit \glspl{adc} are involved due to distinct dominant degrading factors at low and high \gls{snr}, i.e., the \gls{awgn} and the 1-bit quantization distortion, respectively. Specifically, at high \gls{snr}, the 1-bit quantization distortion makes the soft-estimated symbols corresponding to transmit symbols with the same phase indistinguishable, leading to high \gls{ser} \cite{Jac17,Atz22,Atz21a}. Therefore, with 1-bit \glspl{adc}, the \gls{ser} performance is no longer monotonic with the \gls{ue} transmit power, and existing uplink \gls{pc} techniques are not able to tackle this behavior.

In this paper, we fill this gap by proposing uplink \gls{pc} methods for massive \gls{mimo} with 1-bit \glspl{adc}. Considering a single-\gls{ue} scenario, our objective is to tune the \gls{ue} transmit power to find the ``right'' operating \gls{snr}, i.e., where the data detection performance is not dominated by either the \gls{awgn} or the 1-bit quantization distortion. To this end, we introduce a multi-amplitude pilot sequence that is utilized at the \gls{bs} to derive the \gls{ue} transmit power adjustments. Then, we present two methods for calculating the \gls{ue} transmit power adjustments using closed-loop uplink \gls{pc}. In the first one, referred to as the single-shot method, the \gls{ue} achieves a highly accurate transmit power level by means of a one-time multi-bit feedback from the \gls{bs}. The second one, referred to as the \gls{dpc} method, employs a continuous 1-bit feedback loop. The latter approach achieves substantial gains epecially in dynamic environments, despite the imprecise power adjustments compared with the single-shot method. Numerical results validate the superiority of the proposed methods compared to conventional approaches.

\vspace{-4mm}

\section{System Model} \label{sec:SM} \vspace{-4mm}

\ifnum\showguides=1
\red{
\begin{itemize}
    \item 1-bit \gls{adc} system model
    \item Signaling/notations
\end{itemize}
}
\fi
We consider a \gls{bs} equipped with $M$ antennas serving a single \gls{ue} equipped with a single antenna in the uplink. Each \gls{bs} antenna is connected to two 1-bit \glspl{adc} for the in-phase and the quadrature components of the input signal. The received signal at the \gls{bs} prior to the 1-bit \glspl{adc} is given by
\vspace{-1mm}
\begin{equation}
\Y = \sqrt {\rho } \h \x^\herm + \Z \in \Compl^{M\times N},
\vspace{-1mm}
\end{equation}
where $\rho$ is the \gls{ue} transmit power, $\h \in \Compl^{M \times 1}$ is the uplink channel, $\x \in \Compl^{N\times1}$ is the transmit symbol vector with $\Exp[ \| \x \|^{2} ] = N$, and $\Z \in \Compl^{M \times N}$ is the \gls{awgn} with entries distributed as $\setC\setN(0,1)$. As in \cite{Jac17,Atz22,Atz21a}, we assume that the channel is subject to i.i.d. Rayleigh fading, i.e., $\h \sim \setC\setN(0,\I_{M})$. However, the proposed framework applies to correlated channel models as well. 
Let us introduce the 1-bit quantization function $Q:\Compl \to \setQ$, with ${Q(a) = \sgn\big(\Re[a]\big) + j\sgn\big(\Im[a]\big)}$
and $\setQ = \{\pm 1 \pm j\}$. At the output of the 1-bit ADCs, we have
\begin{align}
    \R = Q(\Y) \in \setQ^{M\times N}.
\label{eq:quantized-rec}
\vspace{-1mm}
\end{align}
In the following, we briefly discuss the channel estimation and the data detection with 1-bit \glspl{adc}.

\vspace{-3mm}
\subsection{Channel Estimation and Data Detection} \label{subsec:CE}

\ifnum\showguides=1
\begin{itemize}
    \item \red{Write about channel estimation using pilots}
\end{itemize}
\fi

\vspace{-2mm}

To estimate the channel, the \gls{ue} transmits a pilot sequence $\p \in \Compl^{\tau \times 1}$ to the BS, with $\norm{\p}^2=\tau$. Consequently, the received pilot signal at the \gls{bs} prior to the 1-bit ADCs is given by
\vspace{-1mm}
\begin{equation}
\Y_{\rmp(\rho)} = \sqrt {\rho } \h\p^\herm + \Z_\rmp \in \Compl^{M\times\tau},
\label{eq:y-p}
\vspace{-1mm}
\end{equation}
where $\Z_\rmp \in \Compl^{M\times\tau}$ is the \gls{awgn} with entries distributed as $\setC\setN(0, 1)$.
At the output of the 1-bit ADCs, we have
\vspace{-1mm}
\begin{equation}
\R_{\rmp(\rho)} = Q(\Y_\rmp) \in \setQ^{M\times\tau}.
\vspace{-1mm}
\end{equation}
Then, the least-squares channel estimate (up to a scaling constant) is obtained by correlating $\R_{\rmp(\rho)}$ with the pilot vector, i.e.,
\vspace{-1mm}
\begin{equation} \label{eq:ch-est}
\hat\h = \R_{\rmp(\rho)} \p.
\vspace{-1mm}
 \end{equation}

Upon estimating the channel, the \gls{bs} obtains a soft estimate of the transmit symbol vector via linear combining as
\vspace{-1mm}
\begin{equation} \label{eq:data-est}
\hat \x = \psi\R^{\herm} \hat{\h} \in \Compl^{N \times 1},
\vspace{-1mm}
\end{equation}
where $\psi$ is a normalization factor such that $||\hat \x||^2 = N$.
Finally, the soft-estimated symbols in~\eqref{eq:data-est} are mapped to one of the transmit symbols, e.g., via minimum distance detection with respect to the expected values of the soft-estimated symbols \cite{Atz22,Atz21a}. 

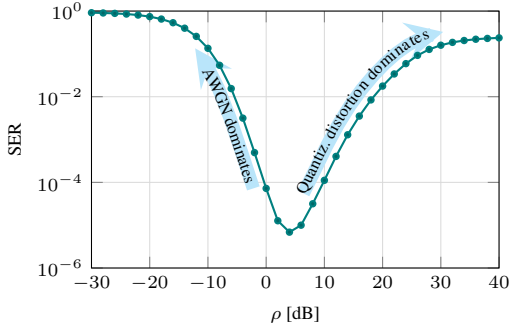
\begin{figure}[t!]
\centering
\begin{tikzpicture}
[auto]

\begin{semilogyaxis}
[
width=7cm,
height=5cm,
xmin=-30, xmax=40,
ymin=10^-6, ymax=10^0,
xtick={-30,-20,-10,0,10,20,30,40},
ytick={10^-6,10^-4,10^-2,10^0},
xlabel = {$\rho$ [dB]},
ylabel = {SER},
ylabel near ticks,
x label style={font=\scriptsize},
y label style={font=\scriptsize},
ticklabel style={font=\scriptsize},
legend style = {font=\scriptsize},
grid=both,
major grid style={line width=.2pt,draw=gray!30},
every axis plot/.append style={thick},
mark options = {solid},
mark size = 1pt,
cycle list name = exotic,
cycle list shift = -4,
]
\addplot [stealth-,vijaya={10}{15}{2.2}{180}]				table [ y=ser, x=snr ] {data/maps.txt};
\addplot [tissa={AWGN dominates}{-1.7ex}{.7}{7}{17}]			table [ y=ser, x=snr ] {data/maps.txt};
\addplot [-stealth,vijaya={19}{31}{1.7}{160.71}]				table [ y=ser, x=snr ] {data/maps.txt};
\addplot [tissa={Quantiz. distortion dominates}{1ex}{.7}{17}{33}]	table [ y=ser, x=snr ] {data/maps.txt};
\addplot								table [ y=ser, x=snr ] {data/maps.txt};
\end{semilogyaxis}
\end{tikzpicture}
\vspace{-3mm}
\caption{\gls{ser} versus \gls{ue} transmit power.}
\label{fig:16-qam-ser-vs-rho}
\vspace{-4mm}
\end{figure}

\vspace{-2mm}
\subsection{Data Detection SER Analysis} \label{subsec:ser}
\vspace{-2mm}

Let $\{s_1, \ldots, s_\Lambda\}$ and $\{\hat{s}_1, \ldots, \hat{s}_\Lambda\}$ be the indices of the transmit symbols and of the detected symbols, respectively. Assuming that all the transmit symbols have equal probability, the \gls{ser} is defined as
\vspace{-1mm}
\begin{equation}
\gls{ser} = \sum_{\lambda = 1}^\Lambda\frac1\Lambda \Pr [s_\lambda \ne \hat{s}_\lambda|s_\lambda],
\vspace{-1mm}
\end{equation}
where $\Lambda$ denotes the order of the modulation scheme.

Fig.~\ref{fig:16-qam-ser-vs-rho} illustrates the 16-\gls{qam} \gls{ser} with the \gls{ue} transmit power for $M=256$. The \gls{ser} is non-monotonic with respect to the \gls{ue} transmit power due to the 1-bit quantization at the \gls{bs}, unlike the unquantized system. Similar to the unquantized case, the \gls{ser} is limited by the \gls{awgn} at the low \gls{snr}, and the \gls{ser} improves with increasing the \gls{ue} transmit power. However, after a certain \gls{ue} transmit power, the \gls{ser} starts to degrade due to the dominance of the quantization distortion, resulting in the aforementioned non-monotonicity.   


\section{Pilot-Based Uplink Power Control} \label{sec:METH}

In this section, we propose pilot-based uplink \gls{pc} methods, a single-shot method, and a \gls{dpc} method for massive \gls{mimo} systems with 1-bit \glspl{adc}, which are specifically tailored to address the non-monotonic data detection performance with respect to the \gls{ue} transmit power. In this regard, we first introduce a multi-amplitude pilot sequence and derive the corresponding \gls{mse} of the pilot estimation with 1-bit \glspl{adc}.

\subsection{MSE of the Composite Pilot Estimation} \label{sec:comp-pilot-mse}

The quantization distortion prevails in receiving the multi-amplitude input signal through the 1-bit \glspl{adc}. Therefore, we consider a pilot sequence with two amplitude levels which we call a composite pilot. Let ${\q = [\sqrt{\rho_1} \p^\tran,  \sqrt{\rho_2} \p^\tran]^\tran \in \Compl^{2\tau \times 1}}$ be the composite pilot with power levels $\rho_1$ and ${\rho_2 > \rho_1}$, where ${\p \in \Compl^{\tau \times 1}}$ consists of uni-modulus symbols. Given the transmitted signal $\q$, the received signal at the \gls{bs} prior to quantization is
\vspace{-1mm}
\begin{align}
\Y_{\rmq} &= [\Y_{\rmp(\rho_1)}, \Y_{\rmp(\rho_2)}] \nonumber \\
          &= [\sqrt{\rho_1}\h\p^\herm + \Z_{\rmp(\rho_1)}, \sqrt{\rho_2}\h\p^\herm + \Z_{\rmp(\rho_2)}] \in \Compl^{M\times2\tau}. \label{eq:y-p-1-2}
\vspace{-1mm}
\end{align}
Note that the average power of $\q$ is ${(\rho_1 + \rho_2)/{2}}$ which we call composite pilot power. The received signal after quantization is
\vspace{-1mm}
\begin{align}
\R_\rmq &= [\R_{\rmp(\rho_1)},  \R_{\rmp(\rho_2)}] \nonumber \\
        &= [Q(\Y_{\rmp(\rho_1)}), Q(\Y_{\rmp(\rho_2)})] \in \setQ^{M\times 2\tau}. \label{eq:com-r-p}
\vspace{-1mm}
\end{align}
%
%
%
%
From the quantized received pilot signal $\R_\rmq$, the channel can be
estimated up to a scaling factor as in \eqref{eq:ch-est} (by replacing $\p$ with $\q$), and the average \gls{mse} of estimating the composite pilot can be computed as
\vspace{-1mm}
\begin{equation}
\begin{aligned}
\gls{pilotmse}\bigg(\frac{\rho_1+\rho_2}{2}\bigg) & =   \Exp\left[\norm{\frac{1}{\norm{\R_\rmq^\herm\R_\rmq \q}} \R_\rmq^\herm \R_\rmq \q - \frac{1}{\norm{\q}}\q}^2\right].
\label{eq:pilot-se-orig}
\end{aligned}
\vspace{-1mm}
\end{equation}
which we call composite pilot estimation \gls{mse} or simply \gls{pilotmse}, where the expectation is taken over the channel and the \gls{awgn} realizations depending on the availability.



\pgfplotstableread[col sep=tab]{data/constel.txt}{\consteltable}
        \pgfplotstablegetrowsof{\consteltable}
        \pgfmathtruncatemacro{\N}{\pgfplotsretval-1}

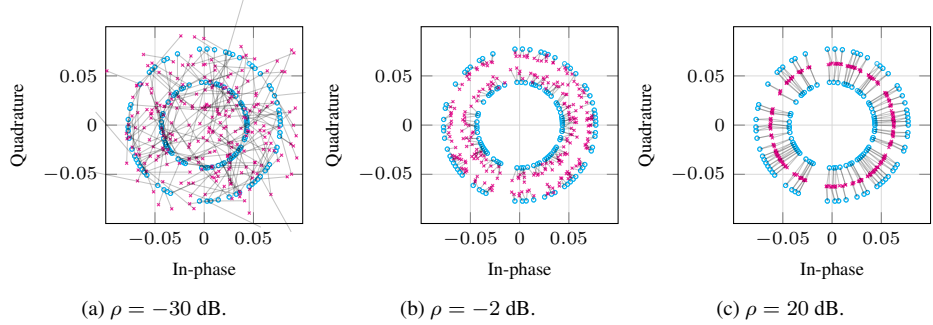
\begin{figure*}[t!]
\centering
\subfloat[$\rho = -30$~dB.]{
\begin{tikzpicture}

\begin{axis}
[
width=4.2cm,
height=4.2cm,
xmin=-.1, xmax=.1,
ymin=-.1, ymax=.1,
xtick={-0.05, 0, 0.05},
xticklabels={$-0.05$, $0$,$0.05$},
ytick={-0.05, 0, 0.05},
yticklabels={$-0.05$, $0$,$0.05$},
xlabel = {In-phase},
ylabel = {Quadrature},
ylabel near ticks,
legend pos = north east,
legend style = {fill opacity = .9, text opacity = 1},
x label style={font=\scriptsize},
y label style={font=\scriptsize},
ticklabel style={font=\scriptsize},
legend style = {font=\scriptsize},
grid=both,
major grid style={line width=.2pt,draw=gray!30},
mark options = {scale=.4, opacity=.8},
cycle list name = exotic,
]

\addplot [mark=o,draw=cyan,only marks]  table [ y=-30qimag,  x=-30qreal  ] {\consteltable}
\foreach \i in {0,...,\N} {
                coordinate [pos=\i/\N] (a\i)
            }; 
\addplot [mark=x,draw=magenta,only marks]  table [ y=-30qhatimag,  x=-30qhatreal  ] {\consteltable}
\foreach \i in {0,...,\N} {
                coordinate [pos=\i/\N] (b\i)
            }; 

\end{axis}
\foreach \i in {0,...,\N} {
            \draw [opacity=.23] (a\i) -- (b\i);
        }
\end{tikzpicture}
\label{fig:pildis--30}
}
\subfloat[$\rho = -2$~dB.]{
\begin{tikzpicture}

\begin{axis}
[
width=4.2cm,
height=4.2cm,
xmin=-.1, xmax=.1,
ymin=-.1, ymax=.1,
xtick={-0.05, 0, 0.05},
xticklabels={$-0.05$, $0$,$0.05$},
ytick={-0.05, 0, 0.05},
yticklabels={$-0.05$, $0$,$0.05$},
xlabel = {In-phase},
ylabel = {Quadrature},
ylabel near ticks,
legend pos = north east,
legend style = {fill opacity = .9, text opacity = 1},
x label style={font=\scriptsize},
y label style={font=\scriptsize},
ticklabel style={font=\scriptsize},
legend style = {font=\scriptsize},
grid=both,
major grid style={line width=.2pt,draw=gray!30},
mark options = {scale=.4, opacity=.8},
cycle list name = exotic,
]

\addplot [mark=o,draw=cyan,only marks]  table [ y=-2qimag,  x=-2qreal  ] {\consteltable}
\foreach \i in {0,...,\N} {
                coordinate [pos=\i/\N] (a\i)
            }; 
\addplot [mark=x,draw=magenta,only marks]  table [ y=-2qhatimag,  x=-2qhatreal  ] {\consteltable}
\foreach \i in {0,...,\N} {
                coordinate [pos=\i/\N] (b\i)
            };

\end{axis}
\foreach \i in {0,...,\N} {
            \draw [opacity=.23] (a\i) -- (b\i);
        }
\end{tikzpicture}
\label{fig:pildis--2}
}
\subfloat[$\rho = 20$~dB.]{
\begin{tikzpicture}

\begin{axis}
[
width=4.2cm,
height=4.2cm,
xmin=-.1, xmax=.1,
ymin=-.1, ymax=.1,
xtick={-0.05, 0, 0.05},
xticklabels={$-0.05$, $0$,$0.05$},
ytick={-0.05, 0, 0.05},
yticklabels={$-0.05$, $0$,$0.05$},
xlabel = {In-phase},
ylabel = {Quadrature},
ylabel near ticks,
legend pos = north east,
legend style = {fill opacity = .9, text opacity = 1},
x label style={font=\scriptsize},
y label style={font=\scriptsize},
ticklabel style={font=\scriptsize},
legend style = {font=\scriptsize},
grid=both,
major grid style={line width=.2pt,draw=gray!30},
mark options = {scale=.4, opacity=.8},
cycle list name = exotic,
]

\addplot [mark=o,draw=cyan,only marks]  table [ y=20qimag,  x=20qreal  ] {\consteltable}
\foreach \i in {0,...,\N} {
                coordinate [pos=\i/\N] (a\i)
            }; 
\addplot [mark=x,draw=magenta,only marks]  table [ y=20qhatimag,  x=20qhatreal  ] {\consteltable}
\foreach \i in {0,...,\N} {
                coordinate [pos=\i/\N] (b\i)
            }; 

\end{axis}
\foreach \i in {0,...,\N} {
            \draw [opacity=.23] (a\i) -- (b\i);
        }
\end{tikzpicture}
\label{fig:pildis-20}
}

\caption{Impact of \gls{awgn} and quantization distortion on the received pilot symbols. \textcolor{cyan}{$\circ$} - transmitted pilot symbols, \textcolor{magenta}{$\times$} - received pilot symbols. } \label{fig:pildis}
\vspace{-3mm}
\end{figure*}

\begin{figure}[t!]
\centering
\begin{tikzpicture}
[auto]
\begin{semilogyaxis}
[
width=7cm,
height=5cm,
xmin=-30, xmax=35,
ymin=1.5 * 10^-2, ymax=.6,
xtick={-30,-20,-10,0,10,20,30,40},
ytick={10^-2,10^-1,10^0},
xlabel = {$\rho$ [dB]},
ylabel = {MSE of the pilot estimation},
ylabel near ticks,
x label style={font=\scriptsize},
y label style={font=\scriptsize},
ticklabel style={font=\scriptsize},
legend style = {font=\scriptsize},
grid=both,
major grid style={line width=.2pt,draw=gray!30},
every axis plot/.append style={thick},
mark options = {solid},
mark size = 1pt,
cycle list name = exotic,
cycle list shift = -4,
]

\addplot [stealth-,vijaya={5}{11}{2.1}{-180}]				table [ y=pilotmse, x=snr ] {data/maps.txt};
\addplot [tissa={AWGN dominates}{-1.7ex}{.7}{2}{14}]			table [ y=pilotmse, x=snr ] {data/maps.txt};
\addplot [-stealth,vijaya={16}{33}{.9}{163}]				table [ y=pilotmse, x=snr ] {data/maps.txt};
\addplot [tissa={Quantiz. distortion dominates}{1ex}{.7}{15}{38}]	table [ y=pilotmse, x=snr ] {data/maps.txt};
\addplot  table [ y=pilotmse, x=snr ] {data/maps.txt};
\end{semilogyaxis}
\end{tikzpicture}
\vspace{-3mm}
\caption{\gls{mse} of the pilot estimation versus \gls{ue} transmit power.} \label{fig:16-qam-approx-data-mse-orig-vs-rho}
\vspace{-4mm}
\end{figure}
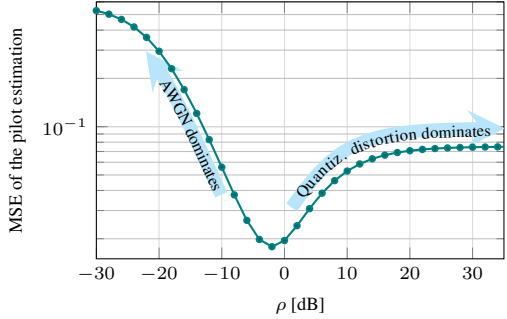

Fig.~\ref{fig:16-qam-approx-data-mse-orig-vs-rho} shows the \gls{pilotmse} versus the composite pilot power ${(\rho_1+\rho_2)/{2}}$  with $M = 256$ and ${\rho_2 - \rho_1 = 5}$~dB. A Zadoff-Chu sequence with $\tau=127$ is used as $\p$. For the above scenario, the level of distortion for 3 interesting $\rho$ values is depicted in Fig.~\ref{fig:pildis} on top of the page. Note that ${\rho_2-\rho_1 = 5}$~dB provides a quantization distortion similar to the 16-\gls{qam} \gls{ser} as shown in Fig.~\ref{fig:16-qam-ser-vs-rho}. Therefore, we can use the \gls{pilotmse} to optimize the \gls{ser} performance over the \gls{ue} transmit power. In the following, we discuss the proposed uplink \gls{pc} methods that utilize the composite pilot estimation \gls{mse}.  


\subsection{Single-Shot Method} \label{sec:ss} 

In open-loop \gls{pc}, the \gls{ue} estimates its pathloss using a downlink broadcast signal called \gls{ssb}. However, there is a \gls{plesterr} error in the pathloss estimate. In the single-shot method, we estimate \gls{plesterr} so that the \gls{ue} transmit power to achieve the \gls{ser} target can be obtained in a single pilot transmission.  To this end, we consider a multi-amplitude pilot. 

In general, for an odd number of amplitude levels $L$, the power of the amplitude level $\ell$ can be obtained (in dB) as
\vspace{-1mm}
\begin{equation}
\rho_\ell = \gls{rhostart}-\gls{gap} + \frac{2(\ell-1)\gls{gap}}{L-1} \quad [\mathrm{dB}], \quad
\ell=1,\ldots, L,
\label{eq:pwr_levels}
\vspace{-1mm}
\end{equation}
where \gls{rhostart} (in dB) is a constant power shift and \gls{gap} (in dB) refers to
the power gap. For example, considering $L=3$, $\gls{rhostart} = 0$~dB, and
$\gls{gap} = 5$~dB, results in the transmitting power levels in $\{-5, 0, 5
\}$~dB. Therefore we can estimate the \gls{pilotmse} with composite pilot with $\{-5, 0\}$~dB and $\{0, 5\}$~dB \gls{ue} transmit power levels, which correspond to the \gls{pilotmse} at {$\rho = -1.8170$~dB and $\rho = 3.1830$~dB} in Fig.~\ref{fig:16-qam-approx-data-mse-orig-vs-rho}, respectively. In this case, the overall span of the \gls{ue} transmit power is $[-5, 5]$~dB.\footnote{It can be shown that the power levels defined in \eqref{eq:pwr_levels} result in $(L+1)/2$ pairs with \gls{gap} difference.}


Let ${\q = \begin{bmatrix}\sqrt{\rho_1}\p^\tran  \ldots \sqrt{\rho_L} \p^\tran\end{bmatrix}^\tran \in \Compl^{ L\tau} \times 1}$ be the transmit pilot sequence with $L$ amplitude levels. We call $\q$ a multi-amplitude pilot. The corresponding received signal after 1-bit quantization is
\vspace{-1mm}
\begin{align}
\R_\rmq &= [\R_{\rmp(\rho_1)}, \ldots, \R_{\rmp(\rho_L)}]  \\
           &= [Q(\Y_{\rmp(\rho_1)}), \ldots, Q(\Y_{\rmp(\rho_L)})] \nonumber\\
           &= [Q(\sqrt{\rho_1}\h\p^\herm\!\!+\!\!\Z_{\rmp_1}), \ldots, Q(\sqrt{\rho_L}\h\p^\herm\!\!+\!\!\Z_{\rmp_L})] \in \setQ^{M\times L\tau}. \nonumber \label{eq:com-r-p_all}
\vspace{-1mm}
\end{align}
We consider any two amplitude level subsequences with a power difference of $\Delta$ out of $L$ subsequences in $\q$ to form the composite pilots. The \gls{pilotmse} values are calculated with~\eqref{eq:pilot-se-orig} having the composite pilot powers of $\gls{compositepowervec} \in \Real^{\frac{L+1}{2} \times 1}$ where $\rho^{\tiny \mathrm{comp}}_i = (\rho_i + \rho_{i+(L-1)/2})/2, i=1, \ldots, (L+1)/2$ (in linear scale). We compute the power offset $\rho_{\mathrm{offset}}$ as
\vspace{-1mm}
\begin{equation}
    \hat{\rho}_{\mathrm{offset}} = \underset{\rho_{\mathrm{offset}}}{\argmin} \|\gls{pilotmse}^{\mathrm{ref}}(\gls{compositepowervec}+\rho_{\mathrm{offset}}) - \gls{pilotmse}^{\mathrm{est}}(\gls{compositepowervec})\|^2,
\label{eq:ss}
\vspace{-1mm}
\end{equation}
where $\gls{pilotmse}^{\mathrm{ref}}(\gls{compositepowervec}+\rho_{\mathrm{offset}})$ is the vector consisting of the reference \gls{pilotmse} values at the composite pilot powers $\gls{compositepowervec}+\rho_{\mathrm{offset}}$ for the given setup, the mapping of which is stored at the \gls{bs}. $\gls{pilotmse}^{\mathrm{est}}(\gls{compositepowervec})$ is the vector consisting of the \gls{pilotmse} estimated values having composite pilot powers of  \gls{compositepowervec}. After computing the $\hat{\rho}_{\mathrm{offset}}$, the \gls{bs} computes the suitable power level to achieve the \gls{ser} target and feeds the power level back to the \gls{ue} using a downlink control channel. This method is useful if the \gls{ue} power is largely deviated from the desired value.

%
%



\vspace{-2mm}
\subsection{DPC Method} \label{sec:grad-meth}
\vspace{-2mm}

Although the \gls{dpc} method is closed-loop, in contrast to the single-shot method, the
\gls{bs} and the \gls{ue} maintain a constant feedback loop in order to keep a desired \gls{dataser}. In addition, the \gls{dpc} method is designed with less pilot and feedback overhead compared to the single-shot method.
In this method, the pilot consists of three amplitude levels having $\rho_1$, $\rho_2$, and $\rho_3$ power levels, such that ${\rho_2 - \rho_1 = \rho_3 - \rho_2 = \Delta}$~dB. Similar to the previous method, the \gls{bs} estimates the \gls{pilotmse} based on~\eqref{eq:pilot-se-orig} with composite pilot powers of ${(\rho_1+\rho_2)/2}$ and ${(\rho_2+\rho_3)/2}$, respectively. Then, the \gls{bs} computes the differential \gls{pilotmse} as
\vspace{-1mm}
\begin{equation}
    \delta_{\gls{pilotmse}} = \gls{pilotmse}^\text{est}\bigg(\frac{\rho_2+\rho_3}{2}\bigg) - \gls{pilotmse}^\text{est}\bigg(\frac{\rho_1+\rho_2}{2}\bigg).
\label{eq:dpc}
\vspace{-1mm}
\end{equation}
If $\delta_{\gls{pilotmse}} > 0$, the \gls{pilotmse} is limited by the quantization distortion at the \gls{bs}, or else by the \gls{awgn}. Given the $\delta_{\gls{pilotmse}}$ value, there exists a one-to-one map between the \gls{pilotmse} and the \gls{dataser}. Subsequently, using the \gls{pilotmse} to \gls{dataser} mapping, the \gls{bs} feeds back the \gls{ue} to increase its transmit power to meet the \gls{dataser} target if the target is not met and the \gls{dataser} is in the \gls{awgn} region similar to the unquantized system. However, increasing the \gls{ue} transmit power to meet the \gls{dataser} in the quantization region leads to degradation in the \gls{dataser}. Therefore, the \gls{bs} feeds back the \gls{ue}  to decrease its transmit power irrespective of the \gls{dataser} target if the \gls{dataser} is in the quantization distortion region. Thus the \gls{dpc} method ensures that the \gls{ue} operates in the \gls{awgn}-limited region at all times. Considering a 1-bit of feedback from the \gls{bs} to the \gls{ue}, the \gls{ue} can only adjust the \gls{ue} transmit power by the step size $\gls{stepsize}$~dB.


\begin{figure*}[!t]
\centering
\includegraphics[width=.65\textwidth]{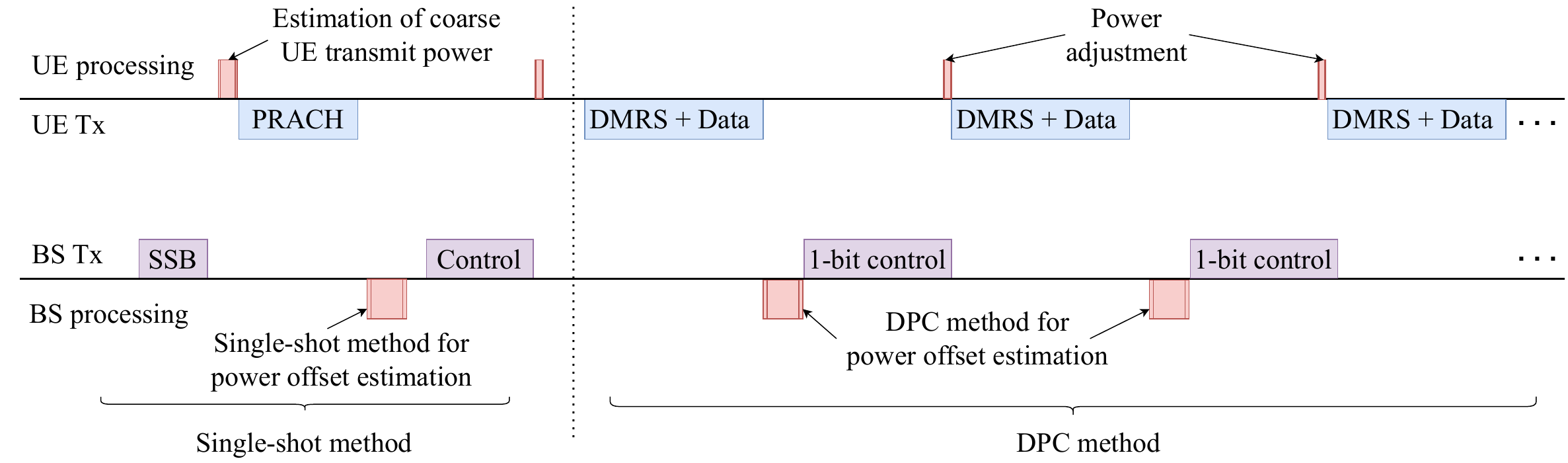}
\vspace{-3mm}
\caption{Signaling and processing flow of the proposed \gls{3gpp} NR-based uplink \gls{pc} implementation.}
\label{fig:3gpp}
\vspace{-4mm}

\end{figure*}

\vspace{-2mm}

 \subsection{3GPP NR-Based Implementation} \label{sec:3gpp}
%
\vspace{-2mm}
 The proposed methods can be integrated with the \gls{3gpp} standard with minimal modifications, as shown in Fig.~\ref{fig:3gpp}. Initially, the \gls{ue} estimate the pathloss from the \gls{ssb} and computes the coarse \gls{ue} transmit power level to achieve the desired performance. Subsequently, the \gls{ue} transmits Zadoff-Chu pilot sequences with $L$ amplitude levels in the \gls{prach} repeatedly with stepwise increase of \gls{rhostart} value until a \gls{rar} is received from the \gls{bs}. This is in contrast to the current standard where a \gls{prach} contains only a single amplitude. In Fig.~\ref{fig:3gpp}, the \gls{bs} happens to detect the \gls{prach} in just one transmission. This allows the \gls{bs} to find the desired power level of the \gls{ue}  using the single-shot method as discussed in Section~\ref{sec:ss}. After computing the desired \gls{ue} transmit power level, the \gls{bs} can give feedback on the power level using a control signal. However, in the connected mode, if the desired power deviation is small due to shadow or large-scale fading, the single-shot method may not be optimal due to resource overhead. Therefore, the  \gls{dpc} method can be utilized to control the small deviations in the power level. In this, the \gls{ue} transmits \gls{dmrs} (DMRS) pilot with three amplitude levels along with the data. Then, the \gls{bs} computes the increment or decrement of the \gls{ue} transmit power in steps of \gls{stepsize} based on the \gls{dataser} requirement as discussed in Section~\ref{sec:grad-meth}. Then, the \gls{bs} provides a 1-bit of feedback to the \gls{ue} using the control signal. Subsequently, the \gls{ue} adjusts its power by $\pm\gls{stepsize}$ in the next transmission.  
\vspace{-2mm}
\section{Numerical Results} \label{sec:numbers}
\vspace{-2mm}
\ifnum\showguides=1
\red{
\begin{itemize}
    \item System parameters, values
    \item At least 2 figures... shows the benefit of proposed \gls{pc} with
some numerical values.
\end{itemize}
}
\fi




We consider $M=256$ \gls{bs} antennas. The power difference in the composite pilot is $\Delta = 5$~dB. 
The \gls{mse} and \gls{ser} reference tables against the \gls{ue} transmit power are generated with $10^6$ Monte-Carlo iterations for each of the simulated $(M, \tau, \gls{gap})$ combination and are stored at the \gls{bs}. Finally, there exists a one-to-one map of \gls{pilotmse} to \gls{dataser} based on the \gls{awgn}-dominant region and the quantization-distortion-dominant region.  

To evaluate the performance of the single-shot method, we assume that the power offset at the \gls{ue}, i.e., \gls{plesterr}, is uniformly distributed in $[-5,5]$~dB. Then, the \gls{ue} transmits the pilot with $L$ amplitude levels and the length of each uni-modulus signal is $\tau$. The overall span of the \gls{ue} transmit power levels is $10$~dB, and the power of each level is chosen as in~\eqref{eq:pwr_levels} where $\gls{rhostart} = 0$~dB. Fig.~\ref{fig:single-shot-all} shows the average squared error in the power offset estimation for different values of $L$ and $\tau$. When $\tau$ is constant, the average squared error in the power offset estimation decreases as $L$ increases. For a given $L$, increasing $\tau$ always gives better performance. When the pilot overhead, i.e., $L\tau$, is constant, the performance degrades as we increase $L$, suggesting that $\tau$ has a stronger impact on the performance compared to $L$.
\begin{figure}
\centering
\begin{tikzpicture}

\begin{axis}
[
width=8cm,
height=4.8cm,
xmin=3, xmax=9,
ymin=0.1, ymax=0.35,
xtick={3,4,5,6,7,8,9},
ytick={0.1,0.15,0.2,0.25,0.3,0.35},
xlabel = {$L$},
ylabel = {\gls{plesterresterrvar}},
ylabel near ticks,
legend pos = north east,
legend style = {fill opacity = .7, text opacity = 1},
x label style={font=\scriptsize},
y label style={font=\scriptsize},
ticklabel style={font=\scriptsize},
legend style = {font=\scriptsize},
grid=both,
major grid style={line width=.2pt,draw=gray!30},
every axis plot/.append style={thick},
mark options = {solid},
cycle list name = exotic,
cycle list shift = 1,
]

\addplot  table [ y=tau105,  x=L  ] {data/ss.txt}; \addlegendentry{\smaller $\tau = 105$}
\addplot  table [ y=tau63,   x=L  ] {data/ss.txt}; \addlegendentry{\smaller $\tau = 63$}
\addplot  table [ y=tau45,   x=L  ] {data/ss.txt}; \addlegendentry{\smaller $\tau = 45$}
\addplot  table [ y=tau35,   x=L  ] {data/ss.txt}; \addlegendentry{\smaller $\tau = 35$}
\addplot  table [ y=Ltau315, x=L2 ] {data/ss.txt}; \addlegendentry{\smaller $L\tau = 315$}

\end{axis}

\end{tikzpicture}
\vspace{-3mm}
\caption{\gls{plesterresterrvar} versus $L$.
}
\label{fig:single-shot-all}
\vspace{-4mm}
\end{figure}

\begin{figure}
\centering
\begin{tikzpicture}

\begin{semilogyaxis}
[
width=8cm,
height=4.8cm,
xmin=0, xmax=2.5,
ymin=10^-5, ymax=10^0,
xtick={0,0.5,1,1.5,2,2.5},
ytick={10^-5,10^-4,10^-3,10^-2,10^-1,10^0},
xlabel = {$t$~[s]},
ylabel = {SER},
ylabel near ticks,
legend pos = north east,
legend style = {fill opacity = .7, text opacity = 1},
x label style={font=\scriptsize},
y label style={font=\scriptsize},
ticklabel style={font=\scriptsize},
legend style = {font=\scriptsize},
grid=both,
major grid style={line width=.2pt,draw=gray!30},
every axis plot/.append style={thick},
mark repeat = {7},
mark options = {solid},
mark size = 1.2pt,
cycle list name = exotic,
cycle list shift = 2,
]
\addplot table [ y=fixed,        x=t] {data/dpc.txt}; \addlegendentry{\smaller Fixed power}
\addplot table [ y=conventional, x=t] {data/dpc.txt}; \addlegendentry{\smaller Conventional PC}
\addplot table [ y=proposed,     x=t] {data/dpc.txt}; \addlegendentry{\smaller DPC method [proposed]}
\addplot table [ y=target,       x=t] {data/dpc.txt}; \addlegendentry{\smaller SER target}
\end{semilogyaxis}
\end{tikzpicture}
\vspace{-3mm}
\caption{\gls{ser} versus time for the proposed closed-loop uplink \gls{pc} strategies.}
\label{fig:differential-pc-all}
\vspace{-4mm}
\end{figure}
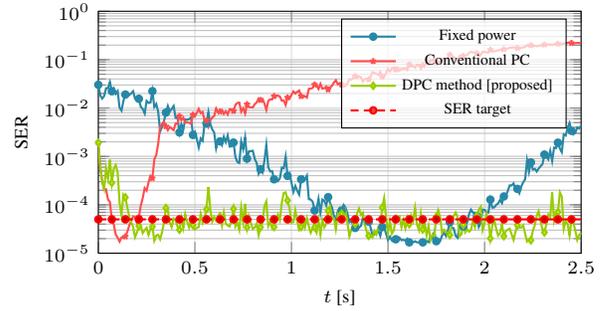

To evaluate the performance of the \textit{\gls{dpc} method}, we consider $\gls{stepsize} = 0.5$~dB,  $L=3$, $\tau=7$, $\gls{rhostart} = 0$~dB, and a shadowing environment as described in~\cite[Ch.~2]{Tse04}, with an \textit{\gls{dataser} target} of ${5\times10^{-5}}$. The pathloss model is ${-61-30\log_{10}(d)}$~dB where $d$ is the distance from the \gls{bs} to the \gls{ue} in meters. At ${t=0}$, the distance between the \gls{ue} and the \gls{bs} is approximately ${63.73}$~m, considering a pathloss compensation with an offset of ${\gls{plesterr} = -5}$~dB based on the \gls{ssb}. The \gls{ue} moves $50$~m with a constant velocity of $20$~m/s towards the \gls{bs} subject to shadowing. The \gls{pc} feedback rate is $100$~bps. The performance of the proposed \textit{\gls{dpc} method} is compared with the \textit{conventional \gls{pc}} method, where the \gls{ue} transmit power is increased if the target \gls{dataser} is not met or else it is decreased by \gls{stepsize}. Also, we compare with the \textit{fixed power}, i.e., \gls{ue} transmits with a fixed power irrespective of the \gls{dataser} target. In Fig. \ref{fig:differential-pc-all}, the 16-\gls{qam} \gls{dataser} is plotted over time, where the shadow and large-scale changes with time. The \gls{dataser} target is met with the proposed \textit{\gls{dpc} method} with a reasonable level of fluctuation. However, using the \textit{conventional \gls{pc}} method may degrade the \gls{dataser} if the \gls{bs} is in the quantization distortion region. Eventually, it can perform worst than the \textit{fixed power} case. The \gls{dataser} of the \textit{fixed power} case changes with the shadow and large-scale fading. Hence, the proposed \textit{\gls{dpc} method} is robust against shadow and large-scale fading.

\vspace{-4mm}
\section{Conclusions} \label{sec:CONCL}
\vspace{-2mm}

This paper explores the uplink \gls{pc} of single-\gls{ue} \gls{mimo} systems with 1-bit \glspl{adc}. The presence of non-monotonic behavior in \gls{mse} or \gls{ser} with respect to the \gls{ue} transmit power with 1-bit \glspl{adc} renders the standard uplink \gls{pc} approach less effective in optimizing \gls{ue} transmit powers for improved performance. To address this challenge, we propose a novel approach that leverages a multi-amplitude pilot received at the \gls{bs} to tune the \gls{ue} transmit power using the single-shot and the \gls{dpc} methods. The single-shot method determines the desired power level of the \gls{ue}, using only one pilot transmission stage. Next, we introduce the \gls{dpc} method, where the \gls{ue} transmit power is adjusted incrementally with a fixed step size, guided by the \gls{ser} behavior mapped from the \gls{mse} of the received pilot signals at the \gls{bs}. To evaluate the effectiveness of the proposed algorithm, we conducted experiments in an environment with shadow fading. The results demonstrate the superiority of the proposed methods over standard approaches. Future work will consider extensions to multi-\gls{ue} settings involving more realistic channel models.


\bibliographystyle{IEEEtran}
\bibliography{IEEEabbr,refs}
\end{document}